# A hybrid single quantum dot coupled cavity on a CMOS-compatible SiC photonic chip for Purcell-enhanced deterministic single-photon emission

Yifan Zhu,[1,2,*] Runze Liu,[3,*] Ailun Yi,[1,2,*] Xudong Wang,[1,2] Yuanhao Qin,[1,2] Zihao Zhao,[1,2] Junyi Zhao,[4,5] Bowen Chen,[1,2] Xiuqi Zhang,[1,2] Sannian Song,[1,2] Yongheng Huo,[4,5,†] Xin Ou,[1,2,‡] and Jiaxiang Zhang[1,2,§]

1. State Key Laboratory of Materials for Integrated Circuits, Shanghai Institute of Microsystem and Information Technology, Chinese Academy of Sciences, 865 Changning Road, Shanghai, 200050, China
2. Center of Materials Science and Optoelectronics Engineering, University of Chinese Academy of Sciences, Beijing, 100049, China
3. Department of Physics, The Chinese University of Hong Kong, Shatin, New Terrotories, Hong Kong, 999077, China
4. Hefei National Research Center for Physical Sciences at the Microscale and School of physical Sciences, University of Science and Technology of China, Hefei, 230026, China
5. Shanghai Research Center for Quantum Science and CAS Center for Excellence in Quantum Information and Quantum Physics, University of Science and Technology of China, Shanghai, 201315, China

\* These authors contributed equally to this work.
†, ‡, § Corresponding authors: yongheng@ustc.edu.cn, ouxin@mail.sim.ac.cn, jiaxiang.zhang@mail.sim.ac.cn.

## Abstract

The ability to control nonclassical light emission from a single quantum emitter by an integrated cavity may unleash new perspectives for integrated photonic quantum applications. However, coupling a single quantum emitter to cavity within photonic circuitry towards creation of the Purcell-enhanced single-photon emission is elusive due to the complexity of integrating active devices in low-loss photonic circuits. Here we demonstrate a hybrid micro-ring resonator (HMRR) coupled with self-assembled quantum dots (QDs) for cavity-enhanced deterministic single-photon emission. The HMRR cavity supports whispering-gallery modes with quality factors up to $7.8 \times 10^3$. By further introducing a micro-heater, we show that the photon emission of QDs can be locally and dynamically tuned over one free spectral ranges of the HMRR (~4 nm). This allows precise tuning of individual QDs in resonance with the cavity modes, thereby enhancing single-photon emission with a Purcell factor of about 4.9. Our results on the hybrid integrated cavities coupled with two-level quantum emitters emerge as promising devices for chip-based scalable photonic quantum applications.

## Introduction

Quantum integrated photonic circuits (QIPCs) play a vital role in quantum information science. It is poised to deliver notable benefits to implement photonic quantum technologies by allowing dense integration of optical components in a single chip with significant stability and reliability[1,2]. Since their initial proof-of-concept demonstrations[3,4], QIPCs have undergone rapid development, with special efforts devoted to silicon-based material platforms like $Si_3N_4$ and silicon[5]. These platforms offer cost-effective manufacturing processes and enable a wide range of photonic quantum technologies such as high-dimensional entanglement[6], quantum communication[7,8], and photonic quantum computation[9,10]. Despite their maturity, limitations have been seen due to the lack of direct bandgap and intrinsic electro-optic effect. The demand for higher levels of integration of active devices, such as single-photon sources (SPSs) and high-speed light modulators, keeps searching for suitable material platforms for the advanced development of QIPCs.

Achieving the goal of integrating both passive and active photonic devices has recently prompted significant research on integrated photonics based on direct bandgap semiconductors. The use of direct bandgap semiconducting materials is advantageous because they can provide various functionalities that are not available in silicon. Among these materials currently being investigated, silicon carbide (SiC) holds great promise due to its desirable optical properties[11,12], including large nonlinearities, CMOS-compatible processing, high refractive index for light confinement, and a wide transparent window spanning near- to mid-infrared wavelengths. Additionally, taking advantage of its intrinsic electro-optic effect, SiC enables the direct integration of fast light modulators based on resonator filter configurations[13]. The advancements in developing integrated photonics with SiC, particularly the 4H-SiC polytype, are remarkable. The 4H-SiC platform has excellent crystalline quality and minimal intrinsic material absorption loss, enabling the realization of wafer-scale materials[14] and the development of high-quality nonlinear photonic devices[15-17]. The compatibility of thin-film 4H-SiC with industry-standard nanophotonics processing technique, together with complex device functionalities, makes the 4H-SiC platform a leading candidate for foundry-scale quantum photonics, with fully integrated, passive and active photonic devices[11].

Although 4H-SiC semiconductor has shown great promise, the breakthrough of advanced QPICs is still hindered by the challenge of integrating efficient SPSs on the chip. To circumvent this limitation, III-V compounds, especially the low-dimensional semiconductor quantum dots (QDs), stand out because of the potential for nonclassical light emission. Unlike nonlinear quantum light sources based on parametric down-conversion or spontaneous four-wave-mixing, self-assembled QDs are well known as atomic-like quantum emitters capable of generating deterministic single photons with near-unity quantum efficiency[18-20]. Of interest is that self-assembled QDs are solid-state materials and can be easily embedded into cavities so as to realize all-solid-state quantum electrodynamics (QED) systems[21-23]. Such intriguing QDs-coupled QED systems provide a prominent means to improve key features of single photons, like extraction efficiency[24-26] and photon indistinguishability[27-29]. Moreover, the solid-state characteristic lends themselves to on-chip integration, promising future scalability. Until now, self-assembled QDs have already been integrated into microring[30] and photonic crystal cavities[31,32] on GaAs photonic chips. However, GaAs is susceptible to high propagation loss and thus is not suited for large-scale integrated photonics. As an alternative approach, hybrid QPICs have opened the possibility to address the complications. Leveraging existing semiconductor processing technology, the hybrid strategy enables a seamless integration of a full set of photonic devices made of distinct materials in a single chip, thus providing a viable solution to enhance the circuit complexity, functionality and compatibility[33,34]. Recent attempts have been made to integrate QDs-coupled cavities on silicon[35,36] and silicon nitride[37,38] photonic chips in a hybrid approach, representing a progress in this direction. However, challenges related to wafer-scale material bonding, substrate thinning, and sophisticated electron-beam patterning hinder their widespread adoption. Additionally, the lack of precise local tuning knobs for controlling the emission properties of individual QDs with respect to cavity modes poses a challenge for scaling up photonic quantum processors. While hybrid integration of QDs-coupled cavities on 4H-SiC would be practically favorable. Combining both semiconducting materials in a single chip could lead to advanced QPICs comprising both deterministic SPSs and high-speed electro-optical reconfigurable photonic circuits, paving the way toward scalable and practical photonic quantum applications.

In this letter, we propose and demonstrate a new method to explore a hybrid microcavity on a CMOS-compatible 4H-SiC photonic chip. Our approach involves the use of a hybrid micro-ring resonator (HMRR) coupled with QDs to achieve cavity-enhanced deterministic single-photon emission. The chip-integrated QED system is based on a 4H-SiC racetrack ring resonator with a transferred GaAs waveguide containing self-assembled QDs. Micro-photoluminescence (μ-PL) at cryogenic temperature reveals that the HMRR supports whispering-gallery modes (WGMs) with quality factors (Q) up to $7.8\times10^3$. By further integrating a micro-heater in the vicinity of the HMRR, we have demonstrated the ability to locally and dynamically tune the photon emission of QDs over one free spectral ranges (FSR) of the HMRR cavity (approximately 4 nm with a tuning rate of $0.13 \pm 0.03$ nm mW$^{-1}$). This precise tuning allows us to engineer individual QDs on resonance with the cavity modes, resulting in enhanced single-photon emission with a Purcell factor of about 4.9. Our results on the hybrid integration open up new perspectives for controlling light emission from single quantum emitters within integrated photonic circuits, paving the way for future developments in quantum information processing and communication.

## Results

Fig.1a shows a sketch of the hybrid microcavity. The quantum emitters used here are InGaAs QDs embedded in the middle of a 180 nm thick GaAs layer, which was grown on a 200 nm thick AlGaAs sacrificial layer on GaAs (001) substrate by solid-source molecular beam epitaxy. Nanophotonic waveguides with a cross-sectional size of 400×180 nm was then fabricated on the GaAs layer. Subsequently we bonded the waveguides onto the surface of 4H-SiC racetrack microring resonators by a transfer-printing technique (Supplementary Note 1 and Fig. S1). Mode transformers composed of tapers with the size decreasing from 400 to 80 nm were fabricated at both ends of the GaAs waveguide. This design ensures light coupling between the GaAs waveguide and the underlying 4H-SiC waveguide. In such a way, conditions for light circulation in the hybrid ring geometry are preserved, suggesting the potential support for WGMs. This is verified in a finite difference time domain simulation on the transmission curve as shown in Fig.1b. Clear peaks associated with fundamental TE-like modes in the near infrared range can be seen. For a ring resonator consisting of a curved section with a radius of 15 μm and a straight section with a length of 25 μm, the free spectral range is about 2 nm.

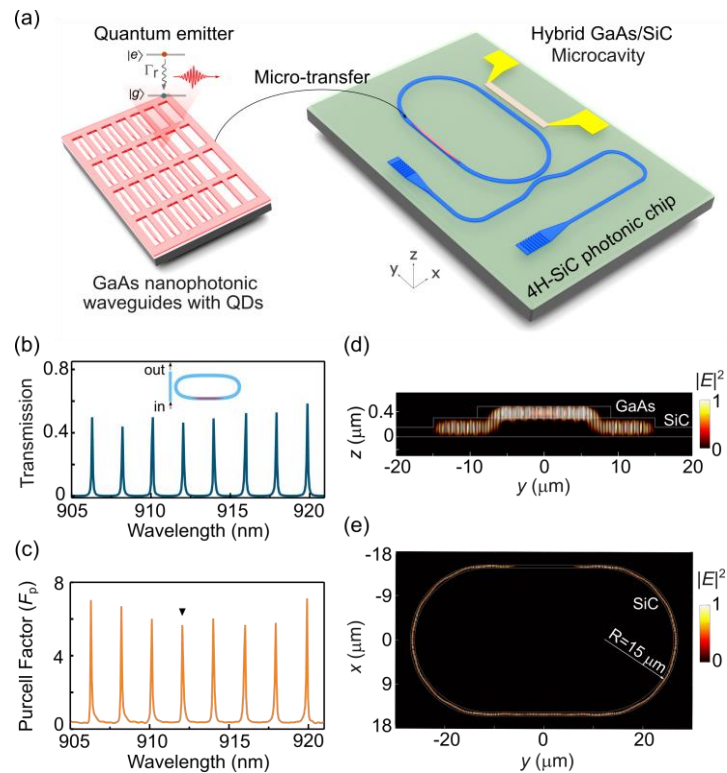

**Fig.1 Schematic illustration of a hybrid GaAs/4H-SiC cavity based on a transfer-printing method. a** QDs containing tapered GaAs and 4H-SiC waveguides are prepared separately by using electron beam lithography and dry etching techniques.

During the transferring, GaAs waveguides are picked up by a PDMS stamp from the GaAs chip and then placed onto a 4H-SiC racetrack resonator to form a hybrid GaAs/4H-SiC cavity. A microheater made of TiN is fabricated to exert thermal-optical tuning for the QDs emission. **b** Simulated transmission spectrum of the hybrid cavity by placing a mode source at one port of the bus waveguide, and monitoring power on the other port. The power is normalized to the maximum value. **c** Calculated Purcell factors of QDs at the center position of the upper GaAs waveguide. Side **d** and top **e** view of the fundamental TE-like mode field of the hybrid cavity when the QD emission wavelength is set in resonant with the WGM peak indicated by the triangular symbol in **c**.

In the presence of the externally transferred GaAs waveguide, additional loss related to the limited coupling of the mode transformer will be added, thereby affecting the quality factors ($Q$) of the HMRR cavity. Such impact can be evaluated by using a simple model: $Q = 2\pi n_g/\lambda(\alpha_{int}+\alpha_{taper})$, where $\alpha_{int}$ and $n_g$ are the intrinsic propagation loss and group index of the 4H-SiC ridge waveguide, $\alpha_{taper}$ represents the coupling loss of the mode transformer. In our previous work, $\alpha_{int}$ was measured to be about 0.78 dB mm$^{-1}$ [39], and $\alpha_{taper}$ depends on the coupling efficiency ($\eta$) of the mode transformer and it can be written as $\alpha_{taper} = -10\times\log_{10}\eta/l$, with $l$ being the length of the mode transformer. With a 10 μm long mode transformer, theoretical calculation yields $\eta$ = 98.3%, resulting in $\alpha_{taper}$ = 7.45 dB mm$^{-1}$ (Supplementary note 2 and Fig. S2). This value implies that $Q$ values of the HMRR will be approximately one order of magnitude smaller than that of the ring resonator without the top GaAs waveguide. Examination of the simulated transmission curve reveals $Q$ values exceeding $1.0\times10^4$ at QDs emission around 910 nm, which is comparable to the previously reported GaAs/SiN hybrid microring resonator[37]. In addition, the theroretical simulation finds out that $\alpha_{taper}$ depends on the geometric size of the GaAs tapered waveguide, and thus can be optimized to improve the $Q$ factor for the hybrid cavity. For quantum dot emission at 910 nm, an optimal $\eta_{taper}$ = 99.8% is found when the GaAs tip width is set to 100 nm and taper length is set to 10 μm (Supplementary note 2 and Fig. S3). With this optimal $\eta_{taper}$, $\alpha_{taper}$ is calculated to be 0.87 dB mm$^{-1}$, approximately one-order of magnitude lower than 7.45 dB mm$^{-1}$. With this low coupling loss for the mode transferormer, the $Q$ factor can be improved by one-order of magnitude and therefore the Purcell factor can also be enhanced accordingly.

When two-level QDs are embedded inside the GaAs waveguide, the HMRR with high-$Q$ resonances can be leveraged to enhance the spontaneous emission rate of QDs through the Purcell effect[30]. For a theoretical estimation, we treat two-level QDs as in-plane electric dipole moments positioned at the center of the GaAs waveguide. The simulated wavelength-dependent Purcell factors of QDs within the cavity are illustrated in Fig.1c, indicating the Purcell factors up to 6 at the HMRR resonances. Fig.1d and Fig.1e show the simulated mode profiles of the WGM as indicated by the black triangle marker in Fig.1c. In stark contrast to previously demonstrated hybrid ring resonators where modes are uniformly confined[30,37], our HMRR features a three-dimensional stack in which modes are confined either in the GaAs or 4H-SiC waveguides and they interact evanescently through the mode transformers.

Based on the theoretical simulations described above, we fabricated the HMRR device utilizing the transfer-printing technique. Fig.2a presents a scanning electron microscope (SEM) image of the device. The HMRR interfaces with a bus waveguide terminated with grating couplers for efficient light collection from the chip to free-space, and vice versa. Fig.2b provides a detailed view of the central region of the hybrid cavity, in which precise integration of the GaAs waveguide onto the surface of the 4H-SiC waveguide is clearly shown. Prior to the GaAs waveguide integration, super-continuum laser light was coupled in the cavity through one of the grating couplers, and the transmission curve of the resonator at room temperature was measured from the other grating coupler, as illustrated in Fig.2c. The resulting resonant peaks indicate a load $Q$ values of approximately $1.8 \times 10^4$.

After transferring the GaAs waveguide, the device was loaded into a closed-cycle cryostat and cooled to approximately ~6 K. A home-made confocal microscope setup was used for optical excitation and collection either from the QD position or grating coupler (Supplementary note 3 and Fig. S4). Optical properties of the hybrid cavity were then examined by exciting the top GaAs waveguide with a high-power continuous-wave green laser (532 nm, 20 µW). This allows observation of saturated PL from QDs via the grating couplers. The spectrum in Fig.2d exhibits regular resonant peaks with the FSR of 2 nm, consistent with the cavity without the top GaAs waveguide. Employing a Lorentzian function to fit one of the cavity modes at approximately $\lambda_c \sim$ 914.7 nm yields a full-width at half-maximum (FWHM) $\Delta\lambda \sim$ 0.117 nm, corresponding to a $Q$ value of 7800. The slight decrease in $Q$ value, compared with the theoretically simulated $Q$ factor, is likely due to the imperfections occurred in the transfer process and additional scattering loss of the waveguides. Next, we decreased the excitation laser power, QDs emission with narrow linewidth was immediately observed as shown in Fig.2e. Based on a polarization-dependent measurement (not shown here), the brightest peak is identified to be the exciton photon (X) and its wavelength (913.6 nm) is spectrally detuned from the nearest cavity mode resonance by ~1.1 nm.

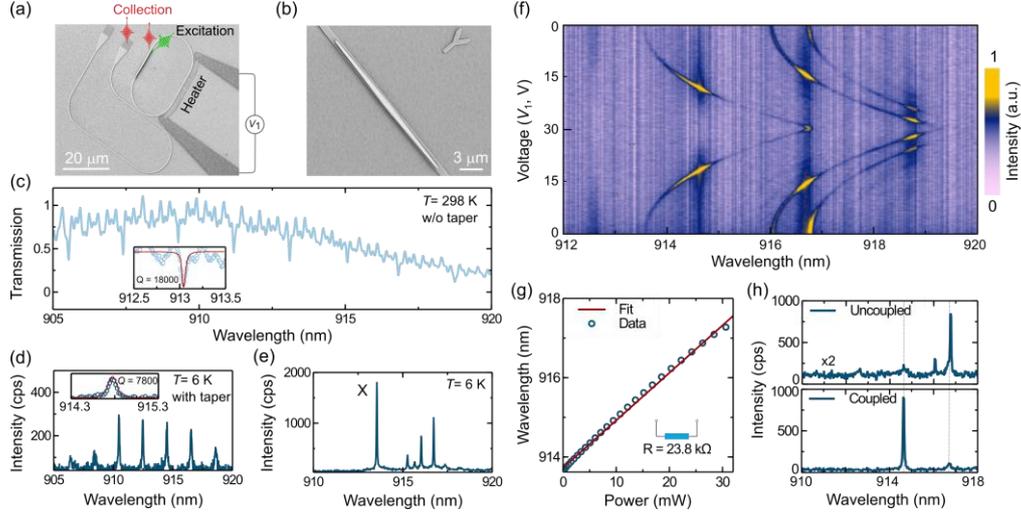

**Fig.2 Experimental fabrication and optical characterizations of the hybrid GaAs/4H-SiC cavity. a** SEM image of the hybrid cavity. QDs inside the upper GaAs waveguide can be non-resonantly excited by a continuum-wave 532 nm laser and the generated light is collected from either the top position of QD or grating couplers. **b** Zoom-in SEM image of the transferred GaAs waveguide on 4H-SiC racetrack ring resonator. **c** Experimentally measured transmission spectrum of a 4H-SiC resonator before transferring the GaAs waveguide. The inset is the zoomed peak at 913.10 nm and the red line is a Lorentzian fit to the resonance. **d** Spectrum of QDs in the upper GaAs waveguide. The QDs were excited with a high-power laser and the distinct peaks are WGMs of the hybrid cavity. **e** Spectrum of QDs excited with a low laser power. The spectrum is collected from the grating coupler. **f** Color plot of μ-PL spectra as a function of power applied to the TiN microheater. The spectra were collected from the grating coupler. **g** Extracted wavelength of the X photon as a function of heating power from 0 to 30 mW. The red line indicates a linear fit to the experimental data. **h** Spectra of the X photon when QD1 is tuned off/on resonance with the cavity, and dashed grey lines corresponds to the adjacent cavity modes.

To dynamically tune the resonance of the QD emission (hereinafter denoted by QD1) with the cavity mode, a TiN microheater was fabricated near the HMRR as shown in Fig.2a. By applying a voltage ($V_1$) ranging from 0 to 30 V, the X photon emission, together with other emission lines, can be dynamically tuned. When the QD emission is tuned to the resonance with the cavity mode, enhanced light emission from the QD is observed, as shown in Fig.2f. A linear dependence of the wavelength of the detected X photons from QD1 on the heater power is found (Fig. 2g). Gien a heater resistance of 23.8 kΩ, a thermal-optical tuning rate $\gamma = 0.13 \pm 0.03$ nm mW$^{-1}$ is extracted. This fast-tuning rate enables a total wavelength shift of about 4 nm, covering a full FSR of the HMRR. With this method, one can expect a possible coupling for all QDs to the HMRR despite the presence of large inhomogeneous broadening (~10 nm) for self-assembled QDs[40]. As the wavelength of the QD1 emission is tuned on resonance with the cavity mode, a pronounced enhancement of photon emission is observed and the relative magnitude of such enhancement is shown in Fig.2h. This enhancement is attributed not only to the increase in emission rate of due to cavity coupling, but also to the photon coupling efficiency from the cavity to the neighboring bus waveguide. To obtain a more quantitative estimate of the photon emission rate due to cavity enhancement, we performed detailed PL lifetime measurements as described below.

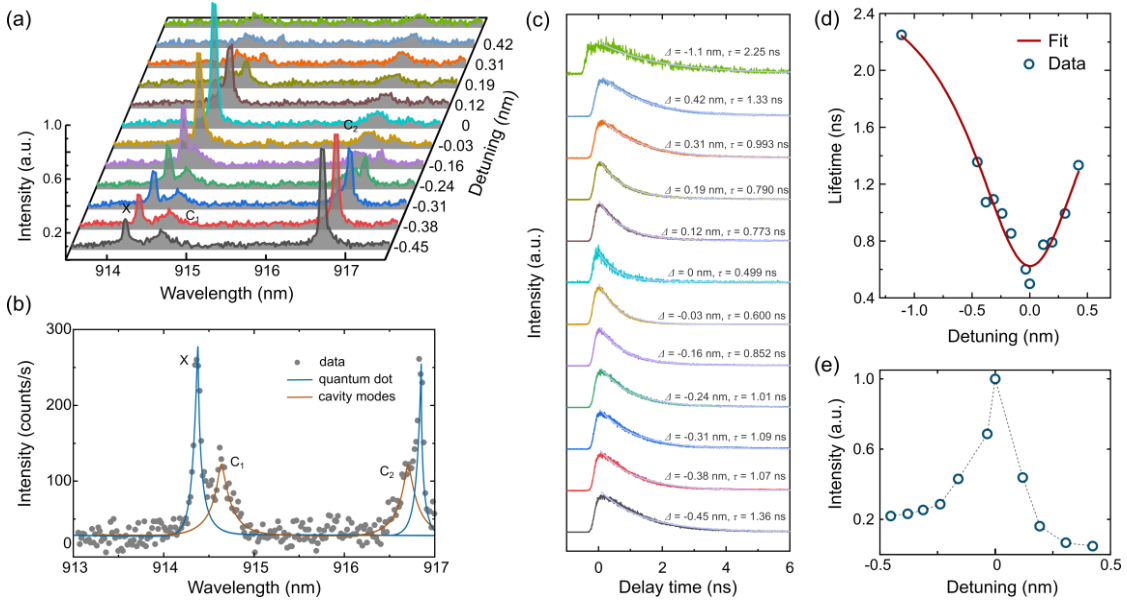

**Fig.3 Time-resolved μ-PL measurements of the X photon of QD1 when tuned off/on resonance with the cavity mode. a** Spectra of the X photon with different wavelength detunings. The spectrum in the top curve (green line) is directly collected on top of QD1 and the others are collected from the grating coupler. **b** PL spectrum at the detuning value of $\Delta$ = -0.24 nm. **c** Corresponding lifetimes ($\tau$) for different wavelength detunings ($\Delta$), the curve color corresponds to the spectrum in **a**. **d** Lifetime as a function of detuning. **e** Normalized intensity as a function of $\Delta$ for the filtered X, obtained from Lorentzian fits as in **a**.

For carrying out time-resolved PL measurement, a picosecond pulsed 532 nm laser with a repetition rate of 40 MHz was used to excite QD1. Fig.3a shows the spectra of the X photon emission ($\lambda_X$) tuned with respect to the cavity mode ($\lambda^1_c \sim$ 914.7 nm for $C_1$ mode) by different detunings ($\Delta = \lambda_X - \lambda^1_c$). A zoomed spectrum in Fig. 3b at the detuning value of $\Delta = -0.24$ nm clearly visualizes both the quantum emission and the cavity modes. In case of far detuning at $\Delta = -1.1$ nm, light coupling from the cavity to the bus waveguide is prohibited. In this scenario, we collected the QD1 spectrum on its top position instead of the grating coupler. As the detuning decreases, the optical coupling efficiency from the cavity to the bus waveguide increases, allowing direct observation of the X photon through the grating coupler. Subsequently we selectively collect the X photon emission by employing a home-built high-resolution grating filter, followed by directing to a time-correlated single photon counting module with which time-tagged single photon arrival events relative to the excitation laser pulses were recorded. Fig.3c displays the lifetime results at different detunings. From the figure, we can see that lifetime decreases as the detuning is tuned from 0.42 to 0 nm, and it increases again as the detuning is varied from 0 to -0.45 nm. The dynamic and reversible changes in the lifetime are consistent with the change of the QD spectral intensity (Fig. 3e), suggesting a pronounced cavity modification for the waveguide-coupled QD emission. We then fit the temporal profiles with multi-exponential decay processes by taking into account of our system response time (100 ps width). The lifetimes of 2.25 ± 0.03 ns at far detuning and 0.499 ± 0.01 ns at zero detuning are obtained. Fig.3d quantitively shows the lifetime as a function of different detuned wavelengths, with the data being well-fitted by a Lorentzian peak function. From the data, we evaluate the Purcell enhancement factor for the QD spontaneous emission rate employing a model presented in Ref.[41,42]. Considering the lifetimes at on-resonance ($\tau_c$) and off-resonace ($\tau_o$), the Purcell factor can be calculated by $F_p = 1+F$, where $F = (\tau_o/\tau_c - 1)/\zeta_{PL}$ ($\zeta_{PL}$ is the total photon emission of the QD1 in the zero-phonon line). For solid-state self-assembled QD systems, $\zeta_{PL}$ is temperature-dependent and typically exceeds 90% at cryogenic temperatures. Therefore, the assessed Purcell enhancement factor is found to be $F_p = 4.9$, a value higher than the previously reported chip-based photonic crystal cavities[30,32,37]. The experimental value is slightly smaller than the theoretically simulated maximum Purcell factor. This discrepancy can be ascribed to the random distribution of the QD position and its dipole orientation with respect to the nanophotonic waveguide (Supplementary note 4 and Fig. S5). In order to achieve an optimal Purcell factor for our hybrid cavity, the well-established QDs positioning technique can be used to place the QD at the center of the waveguide[21].

Thus far we have successfully demonstrated a chip-integrated QDs-coupled QED system. In the next we perform second-order time correlation measurement on this cavity-enhanced QD emission so as to assess its single-photon emission properties. Fig. 4a shows the second-order correlation histogram $G^{(2)}(\tau)$ for the cavity-coupled QD1 under *p*-shell excitation (excitation wavelength at about 902.5 nm). The dip at zero time delay ($\tau = 0$), together with the periodically separated peaks at nonzero time delays, represents deterministic single-photon emission. The multi-photon emission, characterized by $g^{(2)}(0)$, is founded to be as low as 0.008 ± 0.004. Such high purity (99.2%) suggests a superior performance of our device as a promising chip-integrated single-photon source. Aside from the anti-bunching behavior, a pronounced bunching histogram in a long time scale is observed (Fig. 4b). This suggests a blinking effect for the QD single-photon source, which is likely due to the interaction between the bright and other long-lived exciton states (*e.g.* darks states). We determine the time constants of "blinking" into the long-lived exciton states to be 167 ± 7.7 ns and 252 ± 19.6 ns into the bright exciton state, yielding efficiency of 60.1% for QD emission into the bright exciton states.

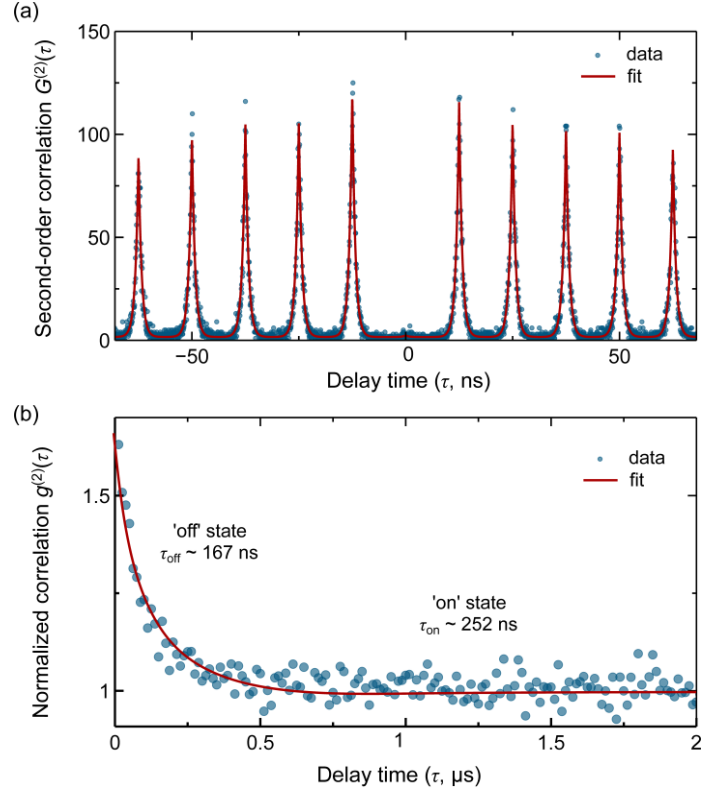

**Fig.4 Single-photon emission properties of the chip-integrated cavity-coupled QD. a** Second-order autocorrelation histogram as a function of delay time for the Purcell enhanced QD1 emission under pulsed quasi-resonant excitation. The red solid lines are theoretical fit convoluted with the instrumental response function (100 ps width). **b** Normalized second-order autocorrelation function $g^{(2)}(\tau)$ in the time scale of 2 μs. The 'bunching' peaks close to the zero time delay indicate a blinking behavior of the source. Two processes, corresponding to a coupling with bright ('on' state) and long-live dark ('off' state) exciton states, are distinguished in the QD single-photon emission.

Empowered by the thermal-optical tuning with the desired dynamical and precise tuning capability, we have so far achieved a chip-integrated deterministic single-photon source with a modest Purcell enhancement . When designing large-scale QIPCs, it is crucial to pursue multimode scalability mediated by cavity-enhanced indistinguishable single-photon sources. This has motivated numerous developments in spectral tuning for both cavities and quantum emitters[43-45]. Although progress has been achieved in stand-alone cavity-coupled QDs devices, developing chip-integrated devices with identical emission wavelengths remains an elusive task. Here, we demonstrate the potential of our device to overcome this obstacle by realizing a local tuning capability. To this end, two HMRRs were fabricated and separated from each other by about 200 μm, as shown in Fig.5a. These two HMRRs were designed to have same geometric size in order for preparing cavity modes with identical resonant modes. Fig.5a shows a microscopic image of the devices, together with the SEM images showing the integrated GaAs waveguides on each HMRR. Microheaters close to each HMRR device were fabricated for thermal-optical tuning. Since they are distantly separated, thermal crosstalk can be suppressed or cancelled. With this design, we carried out spectral tuning for two individual QDs in the two HMRRs by applying voltages $V_2$ and $V_3$ to the corresponding microheater. Fig.5b shows the tuning results and two QDs are tuned to have the same wavelengths at 912.41 nm. When closely looking at the spectrum of each dot (Fig.5c and Fig.5d), they have different wavelengths at 911.4 nm and 912.41 nm, respectively. Cavity modes of the two HMRRs are also visible in the spectrum, and they locate at the wavelength of 912.37 nm and 912.38 nm at cryogenic temperature. We attribute this slight difference in the mode resonance to the fabrication imperfections of the nanophotonic structures. Upon resonance, the time-resolved PL of the two dots were measured and lifetimes of 1.09 ns and 648 ps were extracted, as shown by the inset images. Their Purcell factors are found to be 1.9 and 2.2 respectively. Finally, we verified the cavity-enhanced single-photon emission by measuring the two-photon autocorrelation function $g^{(2)}(\tau)$, as shown in Fig.5e and Fig.5f. Strong anti-bunching at zero time delay is observed. A fit to the data using an effective two-level model revealed $g^{(2)}(\tau) = 0.09 \pm 0.01$ and $0.03 \pm 0.01$ respectively, confirming cavity-enhanced single-photon emission.

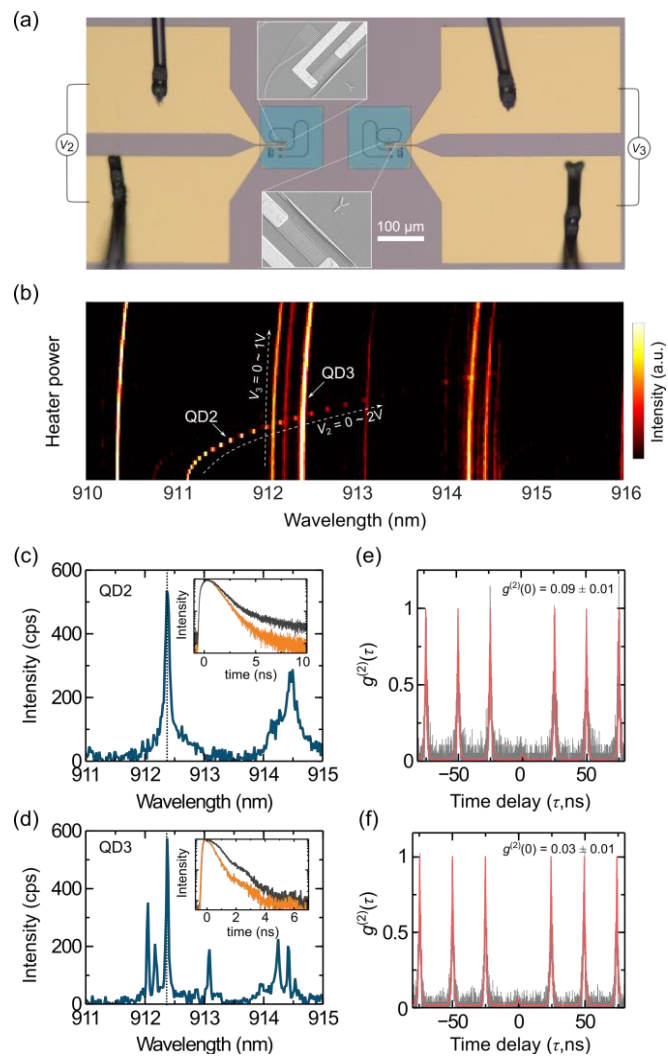

**Fig.5 On-chip local tuning of two distant cavity-coupled QD emitters. a** A microscopic image of two HMRR devices which are separated by about 200 μm. SEM images in insets show the transferred tapered GaAs waveguides. **b** Independent thermal-optical tuning of QDs emission. Exciton emissions of two cavity-coupled QDs on the same chip are controllably tuned into spectral resonance. **c** and **d** PL spectra of QD2 and QD3 when the tuned exciton photon emission is on resonance with the cavity mode. The insets are lifetime measurements. **e** and **f** Normalized second-order autocorrelation functions $g^{(2)}(\tau)$ measured for the QD2 and QD3 when their wavelengths are tuned to the same value.

## Discussion

In conclusion, we have proposed a viable approach to achieve controlled integration of active nanophotonic waveguide onto low-loss photonic circuits through a transfer-printing technique. This innovative approach facilitates the successful implementation of a hybrid single QD-coupled cavity on a CMOS-compatible 4H-SiC photonic chip. Low-temperature micro-PL spectroscopy verified the formation of WGMs with high-quality factors up to $7.8\times10^3$. By further exploring a thermal-optical tuning technique, we have demonstrated a large wavelength tuning range (~4 nm) for the cavity-coupled QDs devices. Such a prominent tuning method allows a dynamic and precise control on QDs emission in resonance with the cavity modes, enabling the demonstration of chip-based cavity-enhanced deterministic single-photon emission with a Purcell factor of 4.9. It is also worth noticing that this large thermal-optical tuning range covers one FSR of the cavity resonances. This unique capability renders coupling of almost all QDs emission to the cavity modes practically convenient, regardless of the presence of large inhomogeneous broadening of self-assembled QDs. At last, we have shown the capability of local tuning of the hybrid cavity systems so that two distant QDs-coupled cavities can be tuned independently to have identical emission wavelengths, a necessary step towards an on-chip quantum network mediated by indistinguishable single photons. As compared to the state-of-the-art works on chip-integrated QDs-based QED system for hybrid quantum photonic circuits (see Table S1), we find that our device features wider tuning range, local tuning capability, higher Purcell factor and purity of deterministic single-photon emission. All these metrics make our chip particularly appealing for the development of scalable chip-intergated quantum light sources. Notably, the integrated solid-state QED systems processed by a simple transfer-printing technique can alleviate the complex fabrication of photonic crystal based integrated cavities, thus constituting an important step toward realization of scalable quantum nodes in large-scale photonic circuits, as well as other hybrid quantum systems for quantum network applications.

Despite the achievements, further works are needed to improve the performance of the device in order to maximize the capability for practical chip-based quantum photonic applications. For instance, the device performance is currently limited by a relatively large mode volume, and thus reducing the ring size and improving the quality factor are highly demanded. To this end, employing an adiabatic tapered ring resonator, which features a small bending radius and high-quality factors at the same time, provides a possible solution to address this task[46]. In addition, active efforts are needed to pursue lifetime-limited single-photon emission from the device. Promising strategy at present is to apply resonant or quasi-resonant excitation[47]. All these improvements present here will offer the possibilities to investigate rich physics in such chip-based all-solid-state QDs-coupled QED systems, thus opening a pathway toward very competitive integrated deterministic cavity-enhanced SPSs for scalable quantum photonic applications.

## Materials and methods

### Device fabrication process

The QDs sample used in our work was grown by molecular beam epitaxy. It consists of a 180 nm thick intrinsic GaAs nanomembrane grown on a 200 nm thick $Al_{0.8}Ga_{0.2}As$ sacrificial layer. QDs locates in the middle of the intrinsic GaAs nanomembrane. Electron-beam lithography (Elionix, ELS-F125G8) was then used to pattern tapered waveguide structures, followed by a dry etching using inductively coupled plasma (ICP, Oxford, Plasma pro 100 Cobra 180) etching on the GaAs layer. Thereafter, the patterned GaAs sample was immersed in diluted hydrofluoric acid solution in order to selectively remove the sacrificial layer. With these nano-fabrications, free-standing GaAs nanophotonic waveguides were obtained and they are ready for deterministic microtransfer in the next step.

In parallel with the GaAs waveguides processing, a 4-inch wafer-scale thin-film 4H-SiC on $Si/SiO_2$ substrate, that is, 4H-SiCOI, was prepared by using ion slicing and direct wafer bonding techniques (see more details in ref.[14,39]). To fabricate low-loss passive photonic circuits, the wafer was then cut into small dies with a size of 1×1 $cm^2$. The small chip was cleaned using RCA procedure, then a 540 nm-thick layer of electron-beam resist (ZEP520A) was spin coated and baked at 180°C. A 100 kV electron-beam lithography (Elionix, ELS-F125G8) was used to define the waveguides and ring resonators with a dose of 220 μC $cm^{-2}$. Subsequently, the device patterns were transferred into the 4H-SiC layer by ICP-RIE (ULVAC, NE-550H) with $SF_6/O_2$. The residual resist was then removed by N-methyl-2-pyrrolidone at 90°C for 3h. The thin 4H-SiC film was etched by 150 nm to form ridge waveguides. The width of the 4H-SiC waveguide was chosen to be 800 nm, which was wider than that of GaAs tapered waveguide. This design ensures a single-mode operation of the hybrid waveguide at QDs emission wavelengths (~910 nm). In the meantime, it provides sufficient space to accommodate the GaAs tapered waveguide to be transferred.

During the micro-heaters preparation, a double layer of photoresist (LOR5A, AZ5214) was spin coated on the top of the patterned 4H-SiC layer. A laser direct writing lithography (Durham Magneto Optics Ltd., Microwriter ML3) was used to define the pattern of the micro-heaters. Thereafter, a solution of tetramethylammonium hydroxide in water (AZ 300 MIF) was used to develop the photoresist. A 80 nm-thick layer of titanium nitride was deposited by sputter deposition (PRO Line PVD 75), and the residual resist was then removed by N-methyl-2-pyrrolidone. The same process was used again to fabricate electrodes (~20 nm Ti/~200 nm Au) for bonding aluminum wires.

With the separately prepared GaAs waveguides and 4H-SiC photonic chip, we further adopted a deterministic transfer-printing technique to fabricate the hybrid GaAs/4H-SiC ring cavity[33,34,48]. The free-standing GaAs tapered waveguides were selectively picked up by a transparent Polydimethylsiloxane (PDMS) stamp. They were aligned relative to the 4H-SiC waveguides under a high-resolution microscope and then brought together until they contacted each other. Subsequently, a tiny force was applied to form a Van der Waals force at the interfaces. In the last step, we peeled off the stamp very slowly to completely release the tapered GaAs waveguides.

**Optical measurement setup**

For optical measurements, the hybrid quantum photonic chip was placed in a closed-cycle (Montana, Fusion F2) and all measurements were carried out at about 6 K. Objective lens (50×, N.A. = 0.65, Mitutoyo) was placed inside the cryostat to improve the system stability. A 3-axis nanopositioner (Nano Precision (Shanghai) Inc.) was used to move the sample with a nanometer precision (~50 nm) and milimeter traveling range (6 mm) at cryogenic temperature. A home-built micro-photoluminescence (μ-PL) setup was used to characterize the hybrid photonic chip. The excitation and collection fibers (single mode fiber, SMF) as well as their associated optics were assembled within a 30 mm cage system and mounted on the top of the cryostat. Optical excitation was performed off-resonantly by a continuous wave 532 nm laser, which was coupled into the μ-PL setup by the objective. The PL signal was coupled into an SMF. With this design, photoluminescence of QDs from either the source position or the grating couplers can be collected by the same objective and then delivered to the collection fiber. By using a home-built grating filter (resolution ~0.1 nm), the photoluminescence of QDs was then selectively filtered and directed to a 750 mm high-resolution spectrometer equipped with a charge-coupled device (CCD).

Time-resolved photoluminescence, as well as photon correlation, were measured by a Hanbury Brown and Twiss (HBT) setup which consists of a non-polarizing 50:50 beam splitter, two identical superconducting nanowire single-photon detectors (SNSPDs, Photon Technology Co., Ltd.), a time-correlated single-photon counting module (Picoharp 300) and other polarization optical components. A pulsed 532 nm laser was used to excite QDs to generate time-dependent single-photon sequences, and synchronized electrical pulse signals. The single-photon signals detected by a SNSPD and the synchronized electrical signal were registered to measure the photoluminescence decay.

**Second-order correlation measurements and fits**

In order to confirm the on-demand emission of single-photon from the HMRR device, the HBT setup was used to measure the histograms of the second-order correlation function $g^{(2)}(\tau)$ of QD emission upon pulsed 532 nm laser. Average value of pulse areas of 20 autocorrelation peaks (except the peak of zero time delay) were extracted and then is used normalized the histogram. The normalized histograms were fitted using the following physical model[30]

$$g^{(2)}(\tau) = \sum_n \exp(-|\tau - n/f|/\tau_{\text{lifetime}})$$

Where $f$ is the repetition rate (40 MHz) of the pulsed 532 nm laser, $\tau_{\text{lifetime}}$ is the recombination lifetime of the studied excitonic emission, $n$ is the number of the autocorrelation peak. The absence of the autocorrelation peak at zero time delay represents the occurrence of the anti-bunching.

## Acknowledgements

The authors would like to thank financial support from National Key R&D Program of China (2022YFA1404604), Science and Technology Commission of Shanghai Municipality (16ZR1442600, 20JC1416200), National Natural Science Foundation of China (Nos. 12074400, U1732268, 62293521, 61874128, 61851406, 11774326 and 11705262), Chinese Academy of Sciences Project for Young Scientists in Basic Research (No. YSBR-112), the Strategic Priority Research Program of the Chinese Academy of Sciences (No. XDB0670303), Shanghai Science and Technology Innovation Action Plan Program (20JC1416200, 22JC1403300), Frontier Science Key Program of Chinese Academy of Sciences (No. QYZDY-SSW-JSC032), Innovation Program for Quantum Science and Technology (No. 2021ZD0300204), Shanghai Municipal Science and Technology Major Project (No. 2019SHZDZX01) , Autonomous deployment project of State Key Laboratory of Materials for Integrated Circuits (No. SKLJC-Z2024-B03) and State Key Laboratory of Advanced Optical Communication Systems and Networks (No. 2024GZKF11).

## Data availability

The data that support the plots within this paper and other findings of this study are available from the corresponding authors upon reasonable request.

## Conflict of interest

The authors declare no competing interests.

## Contributions

Y. H., X. O., and J. Z. conceived the experiments. A. Y. fabricated the 4H-Silicon-carbide-on-insulator material. Y. Z. and A. Y. designed and fabricated the microring fabrication techniques of the 4H-Silicon-carbide-on-insulator. R. L. and Y. H. grew GaAs self-assembled quantum dots by using molecular beam epitaxy. Y. Z. and R. L.

developed the GaAs nanophotonic waveguide. Y. Z. completed the hybrid integration of GaAs nanophotonic waveguide with 4H-SiC microring. Y. Z. and X. W. carried out the device characterizations. Y. Z. and J. Z. processed the experimental data, performed the analysis, and drafted the manuscript. X. O. and J. Z. supervised the project. All the authors contributed to the analysis of the data, discussions, and the production of the manuscript.

*Letters* **22**, 586-593 (2022).